\documentclass[10pt, conference, letterpaper]{IEEEtran}
\pdfoutput=1

\usepackage{booktabs} 
\usepackage{xcolor}
\usepackage{textcomp}
\usepackage{fancyvrb}
\usepackage{setspace}
\usepackage{amssymb}
\usepackage{booktabs}
\usepackage{graphicx}
\usepackage{cite}
\usepackage{url}
\usepackage{scrextend}

\usepackage[tight,footnotesize]{subfigure}

\begin{document}
\title{Scan-and-Pay on Android is Dangerous}

\author{\IEEEauthorblockN{Enis Ulqinaku}
	\IEEEauthorblockA{Department of Computer Science\\ETH Z{\"u}rich, Switzerland\\
		enis.ulqinaku@inf.ethz.ch}
\and
\IEEEauthorblockN{Julinda Stefa}
	\IEEEauthorblockA{Department of Computer Science\\Sapienza University of Rome, Italy\\
		stefa@di.uniroma1.it}
\and	
\IEEEauthorblockN{Alessandro Mei}
	\IEEEauthorblockA{Department of Computer Science\\Sapienza University of Rome, Italy\\
		mei@di.uniroma1.it}}

\maketitle

\begin{abstract}
Mobile payments have increased significantly in the recent years and one-to-one money transfers are offered by a wide variety of smartphone applications. These applications usually support scan-and-pay---a technique that allows a payer to easily scan the destination address of the payment directly from the payee's smartphone screen. This technique is pervasive because it does not require any particular hardware, only the camera, which is present on all modern smartphones. 
However, in this work we show that a malicious application can exploit the overlay feature on Android to compromise the integrity of transactions that make use of the scan-and-pay technique. 
We implement Malview, a proof-of-concept malicious application that runs in the background on the payee's smartphone and show that it succeeds in redirecting payments to a malicious wallet. We analyze the weaknesses of the current defense mechanisms and discuss possible countermeasures against the attack.
\end{abstract}

\section{Introduction}
\label{sec:intro}

\begin{figure}
	\centering
	\includegraphics[width=0.25\textwidth, keepaspectratio]{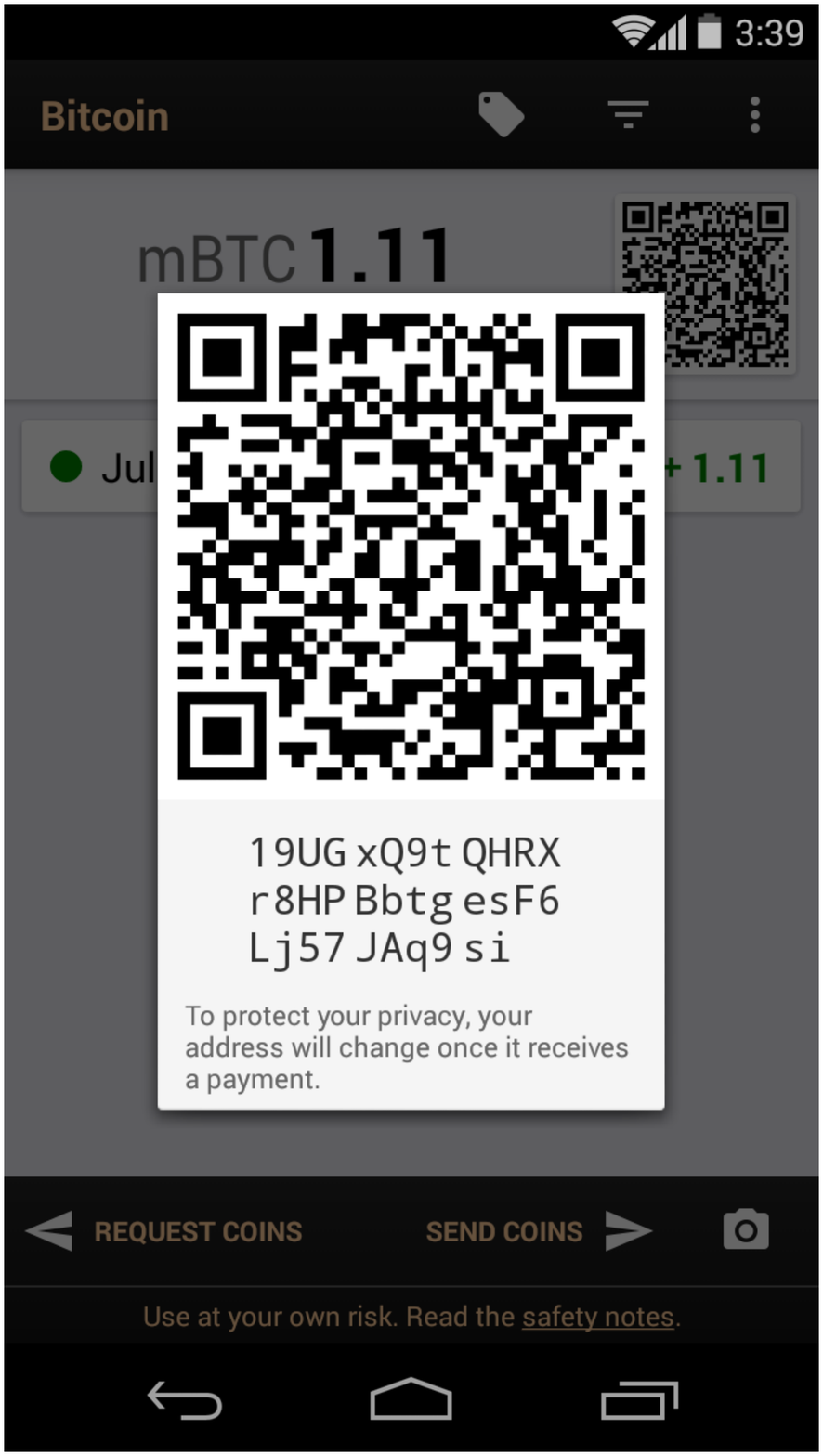}
	\caption{Payee's address QR-code shown on the device screen during a transaction in Bitcoin Wallet.}
	\label{fig:bitcoin-qrc}
\end{figure}

Most mobile applications for one-to-one payments support the scan-and-pay technique---an easy to use method for money transfers. This technique consists on the payer scanning through the mobile's camera the destination address of the payment directly from the screen of the payee's phone. Scan-and-pay does not require any particular hardware, thus it has become very pervasive as it is supported by all modern smartphones. 
It is featured by official applications of traditional banks to enhance user's experience for money transfer. This technique is also popular among applications serving as cryptocurrency wallets.
Differently from traditional banks, cryptocurrencies, such as Bitcoin, rely on cryptography to protect transactions integrity and users' privacy. 
Once a transaction is published to the blockchain it cannot be reverted or modified~\cite{btcDouble}.
Furthermore, as a pair of public and private keys are generated for each transaction, transactions do not give information about the identities of users. 
In general the public key is the user's address and it composed of dozens of case-sensitive random characters (see Figure~\ref{fig:bitcoin-qrc}), which make it difficult for a payer to type it manually into a mobile device. 

The integrity of mobile payments based on scan-and-pay is threatened by malwares that misuse benign features of the system. For instance, Android provides any application with the possibility to show overlays on top of other applications. An overlay can be created by a foreground application or a service---an application component performing long-running operations in the background. The most common ways to create overlays on Android is through \emph{Alert Windows} or \emph{Toasts}. The former requires that the application is granted the \emph{SYSTEM\_ALERT\_WINDOW} permission, while the latter does not require any particular permission. On prior works~\cite{clickjackingBlackhat12, clickjackingUSENIX12, clickjackingUSENIX14, clickjackingSAC15, clickShield} overlays have been studied with particular reference to their potential to mount UI redressing attacks, e.g., clickjacking, or phishing. However, these attacks focus on deceiving the user to commit an unintended action (clickjacking) or steal sensitive information from the user (phishing), while we exploit overlays to compromise the information transferred between two smartphones.

Differently from us, UI redressing attacks employ overlays mostly for tricking the user into clicking an UI component that she does not intend to click~\cite{clickjackingBlackhat12, clickjackingUSENIX12, clickjackingUSENIX14, clickjackingSAC15, vigna15}. For example, a malware places an overlay with the button ``Play'' on top of a sensitive button, i.e., ``Call'', of a victim application. The user sees the button ``Play'' and taps the screen with the intention to start a game, but the overlay is configured in such a way that it does not receive the touch event, therefore the ``Call'' button of victim application is clicked. Although OS offers defense mechanisms, clickjacking still remains an important problem on Android~\cite{clickShield}. 

On the other side, the goal of phishing is to deceive users and steal sensitive information (credentials) from victims, which is orthogonal to our goal.
Chen et. al~\cite{activityAttack} investigates the information available in the \emph{proc} folder to infer the state of the current activity in the foreground and mount phishing attack at the moment the user expects to enter her credentials. Approaches like~\cite{taskJacking15, instantApps} exploit multitasking management systems or instant app features of Android to achieve phishing attacks. 
While previous clickjacking and phishing attacks exploit overlays to deceive the user, we extend the use of malicious overlays to violate the integrity of mobile payments utilizing scan-and-pay technique.

Bianchi et. al~\cite{vigna15} present an exhaustive study on the techniques a malicious application can exploit overlays to trick the user.
They propose an on-device mechanism as a countermeasure that prints on the system navigation bar a secure indicator about the identity of the application on the foreground. Fernandes et al.~\cite{fernandesFC} improve the mechanism proposed in work~\cite{vigna15}. AlJarrah et. al~\cite{maintainUI} introduce a new event \emph{OnOverlap} fired by the system in order to notify the potential victim application about an ongoing attack. Ren et. al~\cite{windowGuard} propose a new security model named Android Window Integrity (AWI) which protects the user against GUI attacks. 
However, most of these defense mechanisms are designed to prevent only UI redressing or phishing attacks and rely on the user to detect malicious overlays. As shown in work~\cite{vigna15}, a significant number of users do not pay attention to security indicators, therefore attacks remain still effective.

In this paper we present an attack that employs the overlay feature in Android to compromise the integrity of one-to-one mobile payments. This attack targets mainly wallet applications for cryptocurrencies such as Bitcoin Wallet~\cite{bitcoinWallet}, among many others.
We do not attack the Blockchain protocol itself but rather the scan-and-pay technique employed by applications that individuals use as cryptocurrency wallets on their smartphones. The attack is significant because it targets a payment method that is very popular and causes financial damage to victims. In the case of cryptocurrencies, the importance of the attack increases because it is impossible to cancel a transaction or get the money back even if the attack is detected by the users immediately after the payment.

To evaluate the efficacy of our attack, we implement \emph{Malview} which runs in the payee's device.
The  malicious behavior of Malview can be hidden as a service of a game or a repackaged popular application~\cite{repackage2}. 
We configure the malicious server on a remote machine that hosts the attacker's Bitcoin wallet. 
Our experiments show that Malview is effective to redirect the payment to a malicious wallet. We show that the attack works well in practice with a phone connected to the Internet using either Wifi or cellular connectivity.

However, the implications of the attack are not limited only to mobile payments. Virtual business card applications are another use-case. These applications employ QR-codes as a simple way to transfer contact details of an individual to another. We describe how Malview can threaten this category of applications and serve as a starting point for more severe attacks against users, e.g., phishing, identity theft, or deploying malicious executable files through emails.

\section{Background: Overlays in Android}
\label{sec:views}
Smartphones are complex systems and the user interface is composed of different components. Android employs the \emph{WindowManager} Interface~\cite{androidDev} to accommodate properly UI components generated by all entities of the system.
\emph{WindowManager} is responsible for managing and generating all the windows, views, buttons, images, and other floating objects on the screen. It is accessible from applications on the foreground or background and it allows them to add or remove overlays on runtime through its methods: \emph{windowmanager.add(view)} and \emph{windowmanager.remove(view)}. 

In the attack presented in this work we make use of views that are placed on top of all the other objects, including the views of the foreground application. These particular view objects can be implemented either as Alert Windows or as Toast Windows~\cite{androidDev}.
Differently from Toast Views, Alert Windows are not restricted by the system for how long they stay on the screen. However, both types of views have a common feature: When created, even from a service on background, Android puts them on top of all other objects, including the foreground application~\cite{androidDev}.
Alert Windows require that the application owns the \emph{SYSTEM{$\_$}ALERT{$\_$}WINDOW} permission in order to create overlays, while toasts do not require any special permission. Interestingly, authors in the work~\cite{cloackDagger} show how a malicious application can misuse Play Store policies to be granted with \emph{SYSTEM{$\_$}ALERT{$\_$}WINDOW} permission even without user approval.
However, the permission is very common among applications on Google Play, as it enriches the user experience on one application while using another one, e.g., a music player widget. Facebook Messenger's ``chat head'' feature is based on the attributes of Alert Windows. 
A search of the Play market through the IzzyOnDroid online crawler reveals that there are more than 600 applications with hundreds of millions of downloads each that require \emph{SYSTEM{$\_$}ALERT{$\_$}WINDOW}~\cite{storecrawler}. Among the very popular applications (with more than a billion downloads) that use the same permission for their features are Facebook for mobile, Skype, the Telegram Messenger, and the Cut the Rope game. The extensive use of this permission by a large number of applications shows that this permission is not perceived as harmful by the users.

\begin{table*}
\centering
\begin{footnotesize}
\caption{Specs of the devices used in the testbed.}
\begin{tabular}{lll}
\toprule
Brand \& OS &CPU &RAM \\ \midrule
Motorola Moto G (Android 4.4.4) & Quad-core 1.2 GHz Cortex-A7 & 1GB\\
Galaxy Note 3 Neo (Android 4.4.2) & Quad-core 1.3 GHz A7 \& dual-core 1.7 GHz A15& 2GB\\
Galaxy Note Pro 12.2 (Android 5.0.2) & 1.9GHz Exynos 5 Octa  & 3GB\\
\bottomrule
\end{tabular}
\label{tab:phones}
\end{footnotesize}
\end{table*}

\section{Attacking Bitcoin Mobile Wallets}
\label{sec:bitcoin}
By default, the application of the payee generates a Bitcoin address of 34 alphanumerical characters associated to the wallet and unique to the transaction---this is done to make the transaction unlinkable to other transactions of the same wallet. The very same feature, however, can be exploited by MalView 
to steal the coins of a transaction: The money leaves the payer, never reaches the payee, and goes to a malicious and anonymous wallet. In terms of usability, a user cannot remember her address because it is not static and changes after each transaction, also it is not easy for a sender to type manually the long random string of the receiver's address, therefore QR-codes offer a soft experience.

All mobile bitcoin wallets employing QR-code based payments operate in a similar way (see Fig~\ref{fig:bitcoin-qrc}). As a proof of concept of our attack, we have implemented it for the Bitcoin Wallet app~\cite{bitcoinWallet}, the most widely used by bitcoiners worldwide (1M--5M installs on Google Play alone) and recommended by the Bitcoin project official homepage \emph{bitcoin.org}.

\section{Adversary Model}
\label{sec:system_model}
We consider a scenario where the attacker controls an app installed on the user's device, similarly to the assumption in~\cite{taskJacking15, activityAttack, tapprints2012, taplogger2012, zhouIdentity13}, with a malicious functionality hidden in a repackege~\cite{repackage2} of a legit application.The goal is to violate the integrity of QR-code based mobile payments without being detected. The malware has access to just two permissions: The \emph{SYSTEM$\_$ALERT$\_$WINDOW}, very common in popular applications, and the \emph{INTERNET} permission---with \emph{PROTECTION$\_$NORMAL} protection level on Android~\cite{androidDev}. This indicates that it is considered not harmful and that it is granted to all apps that require it without asking the user. 

\section{The Attack in Details}
\label{sec:bitcoin-details}
After a mobile wallet application, for example the Bitcoin Wallet, is installed on the device, the application pre-generates the QR-code corresponding to the address to be used in the next transaction. Then, during a QR-based transaction where the sender specifies the amount to be transferred, the parties proceed as follows: The payee loads the QR-code on the device's screen by clicking on the apposite button, pictured as a miniature version of the code (see Figure~\ref{fig:bitcoin-qrc}). The sender uses the app's integrated scanner to scan the payee's QR-code from which the payee's address is extracted. Then, the sender specifies the amount of coins to be sent and taps \emph{Send}, to commit the transaction to the Bitcoin Network for validation~\cite{satoshiBitcoin}. At this point, the payee receives a confirm notification with the transaction details (amount of coins transferred, the payer's address, the time and date of the transaction).

The attack targets users that are the recipients of a transaction, and its goal is to steal the bitcoins destined to the victim's wallet. The idea is to stealthy transform transactions towards the victim into transactions towards another wallet controlled by the attacker; neither the sender nor the receiver notice the attack. In our setup, the Bitcoin Wallet of the attacker executes on a the malicious server. When MalView is installed on the victim's device, it pulls from the server the attacker's address to be used in the next transaction. The attack begins after the target application is started by the victim. Therefore, MalView should be able to correctly detect this event in the victim's device. Malview can utilize known techniques~\cite{vigna15, activityAttack} to detect when the target application is in foreground, in our prototype we make use of \emph{/proc/$<$pid$>$/cgroup} file. 
The malicious application can even trigger system reboot~\cite{forceReboot} to guarantee that the application starts in its main activity. 

When the user starts the Bitcoin Wallet application, MalView exploits the granted permission \emph{SYSTEM$\_$ALERT$\_$WINDOW} to create a transparent malicious overlay on top of the button that loads the legit QR-code on the screen.
At the moment the payee clicks in the application to show her address as a QR-code to the payer, the click goes actually to the Malview. The legitimate application does not receive any event and cannot detect that an attack is taking place. In order for the QR-code to look legitimate to the payer and the payee we use the same code as the legitimate application to generate it. Our target mobile wallet application is open source~\cite{walletSource} and we use the original code to generate the same views that show the payee's address in the victim application. In case that another targeted application is not open source, the adversary can replicate the same design manually also. Therefore, when the unaware victim (the payee) taps the button to load the QR-code of her address on the screen, the event is caught by MalView which loads another QR-code: The one coming from the malicious server. At this point, the transaction that starts from the sender actually arrives on the adversary's wallet. However, in order to keep the attack hidden, the malware on payee's device tries to trick the user that the payment was successful by mimicking notifications of the honest application.

After the transaction is validated by the Bitcoin Network, the attacker's wallet receives the corresponding notification, extracts its details, and sends them to the MalView application---its counterpart on the victim's (payee) device. The Android OS allows any application to show fully customized notifications. MalView exploits this feature and adds the details of the transaction received from the server to create a fake notification on the victim's device. 
This notification is generated on purpose with the icon and the styling of the real Bitcoin Wallet application notification, which makes it indistinguishable from a legit one for the user. As a result, both the sender and the receiver are left unaware of the attack and fooled to believe that the coin transfer went through correctly.

Note that, for the attack to go undetected, MalView needs to know how the Bitcoin Wallet application visualizes the QR-code and the miniature QR-code button on the given victim device. The way an application displays its components on a device is peculiar to the device type, model, and screen resolution, freely accessible on Android through public APIs of the \emph{WindowManager} Interface~\cite{androidDev}.
Then, MalView generates the corresponding parameters regarding the visualization of the Bitcoin Wallet components, and stores them locally in the form of an XML file to be used when needed. 

\begin{figure*}
\centering
	\subfigure[Attack setup time.]{\includegraphics[width=0.25\linewidth, keepaspectratio]{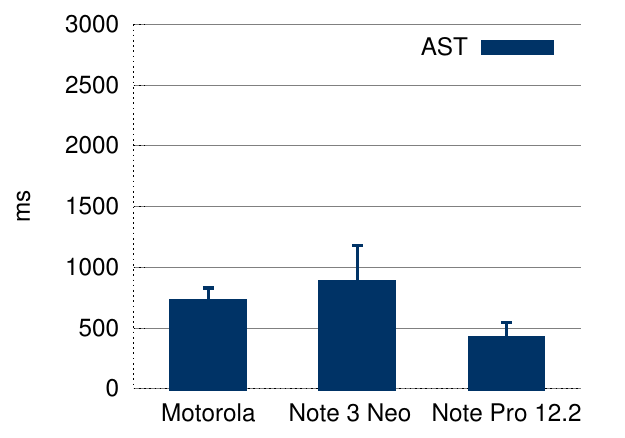}
		\label{fig:bitcoin-efficiency}}\qquad
	\subfigure[Attack responsiveness, WiFi.]{\includegraphics[width=0.25\linewidth, keepaspectratio]{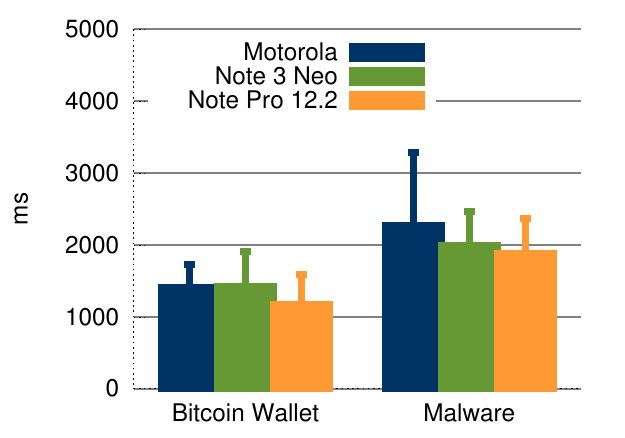}
		\label{fig:wifi-responsiveness}}\qquad
	\subfigure[Attack responsiveness, cellular.]{\includegraphics[width=0.25\linewidth, keepaspectratio]{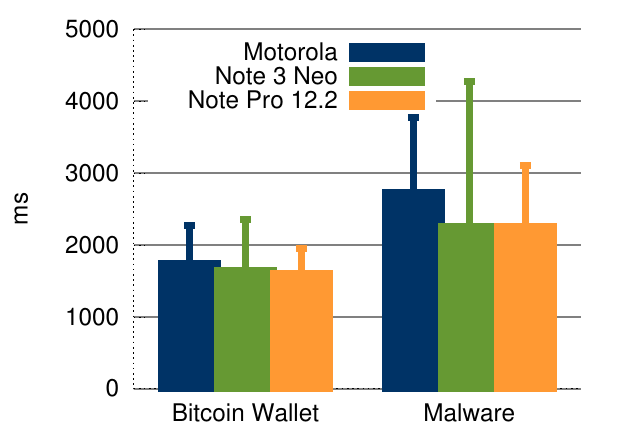}
		\label{fig:4G-responsiveness}}
\caption{Average and standard deviation of the attack setup time and responsiveness (notification arrival time) under different types of connectivity. The setup time is measured from when the user clicks on the Bitcoin Wallet application's icon to when the fake QR-code is shown on screen.}
\label{fig:attack_results}
\end{figure*}

\subsection{Toasts}
\label{sec:toasts}

In our first proof-of-concept implementation of Malvew we exploited the \emph{SYSTEM\_ALERT\_WINDOW} permission to generate the malicious view that shows the fake QR-code on top of the legitimate one. However, a further investigation showed that Malview can be implemented in an alternative way through APIs and functionalities publicly accessible in Android and without requiring the \emph{SYSTEM\_ALERT\_WINDOW} or any other permission at all. Here we describe the details.

Any application in background or foreground can generate notifications or quick messages regarding some aspect of the system through the \emph{Toast} class. The window that shows the volume control while the user is tuning the device's volume is one example of Toast messages. Their purpose is to show some information to the user in a non-intrusive way, therefore they can be employed by any service or user application without requiring any special permission from the user. However, Android does not restrict the design of such messages and one application can fully customize Toasts views to contain images or other objects. Most importantly, Toasts are always shown on top of any other window, including the foreground application, just like Alert Windows. Therefore, an adversary can employ Toasts to develop malicious apps that do not require any special permission from the user as discussed in other works~\cite{clickjackingBlackhat12, vigna15}.

We implemented a version of Malview using Toast Windows instead of Alert Windows as a proof-of-concept application. Given that Android limits the time that a Toast stays on screen to just a few seconds, it becomes problematic for the malware to keep the QR-code on the screen for a longer time. The counterfeit QR-code should be on top of the legitimate one until the payer scans it in order to direct the payment to attacker's wallet. Therefore, Malview exploits the \emph{Toast} class and periodically calls the \emph{toast.show()} method of the malicious view before it expires. In this way the new version of Malview has the same functionalities and performs the same attack without the \emph{SYSTEM\_ALERT\_WINDOW} permission.

\section{The Bitcoin Wallet attack: Experiments}
\label{sec:bitcoin_experiments}
Here we present the results regarding the coin-stealing attack with the Bitcoin Wallet app. The malicious server was deployed on a MacBook Pro machine with a 2.8 GHz Intel Core i5 processor and 8 GB of RAM. For the evaluation we used two Samsung Galaxy Note devices and a Motorola Moto G smartphone (see Table~\ref{tab:phones}). The experiments consisted on sets of 30 QR-code based payments. Each set of payments was performed with and without MalView's presence on the victim (payee's) device. Finally, for the experiments we used both WiFi and cellular connectivity (4G for the two Samsung devices and 3G with the Motorola Moto G device).

\subsection{Attack setup time}
When the victim (payee) opens the Bitcoin Wallet application to start a transaction, MalView starts generating the transparent malicious view that covers the miniature legit QR-code button. This setup phase should complete before the user taps the button to load the QR-code address for the transaction; otherwise, MalView will not be able to capture the tap and load the adversary's QR-code accordingly.
The result presented in Figure~\ref{fig:bitcoin-efficiency} show that MalView is extremely efficient in setting up the attack. Indeed, the average setup time on all devices is less than a second. This guarantees that the malware will seamlessly and stealthily operate to load the adversary's QR-code on the victim's screen on time. 

\subsection{Responsiveness}
After the transaction is submitted by the payer and validated by the Bitcoin Network, the device of the payee receives a notification. When the attack is in place, the actual recipient is the adversary. Thus, the notification is received by the malicious server hosting her wallet. Then, the transaction information is forwarded to the MalView application residing on the victim's device so that it can create and show the fake transaction notification. Clearly, this man-in-the-middle attack incurs some delay in the reception of the notification and, if the delay is high, the users may get alerted. To assess the delay we measure the time interval from when the payer confirms the transaction, till when the fake transaction notification is shown on the victim's device, and compare it with the time interval in a normal transaction. The results are shown in Figure~\ref{fig:attack_results}. We observe that the delay overhead of MalView with respect to the regular application is quite low (around 500 ms).

\section{Other Use-Cases}
\label{sec:otherCases}
The attack described in this paper does not concern only cryptocurrency wallet apps in Android. In fact, it targets a large number of other apps, including e.g. online banking, whose prototype implementation we describe here.

\subsection{Banking Applications}
\label{sec:bankingApps}
\begin{figure}
\centering
\subfigure[The Paytm app. 
]{\includegraphics[width=0.18\textwidth, keepaspectratio]{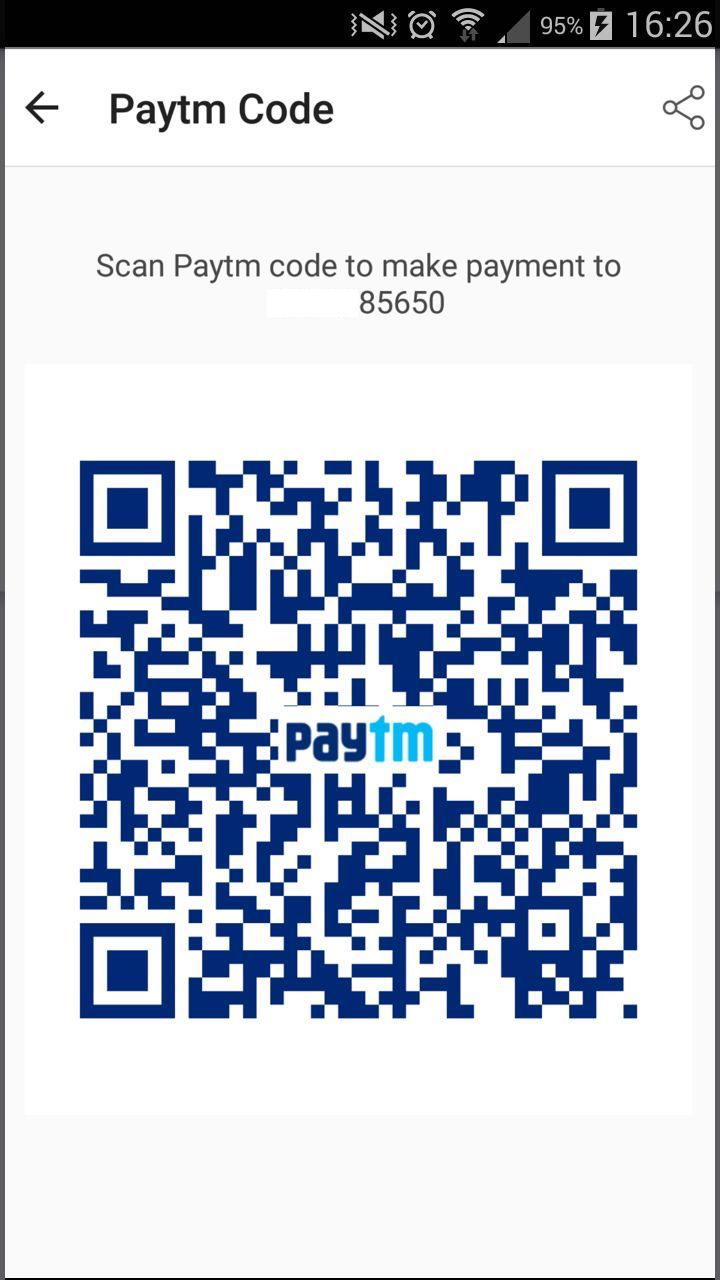}
\label{fig:paytm}}\qquad
\subfigure[The PostePay app.
]{\includegraphics[width=0.18\textwidth, keepaspectratio]{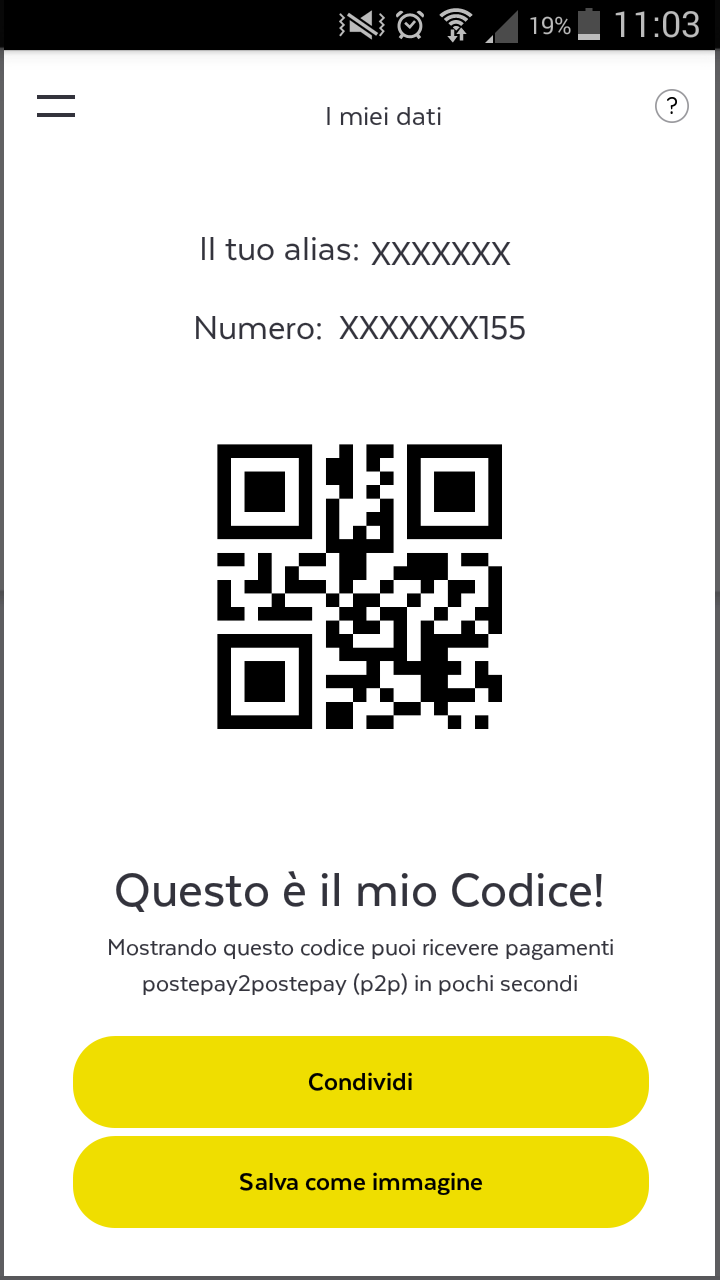}
\label{fig:postepay}}
\caption{Screenshot of two mobile apps employing scan-and-pay through QR-code to read the address of the payee.}
\label{fig:bankapps}
\end{figure}

Mobile payments are spreading rapidly, they are highly pervasive, and countries are moving toward cashless economies. A large currency ban applied by the Indian government forced the indian people to switch to mobile wallets in order to pay for their daily goods and services~\cite{indiaCrise}. Paytm~\cite{paytm} is one very  popular wallet app in India with more than 100M installations. The Paytm app features the scan-and-pay technique in a same way as the Bitcoin wallet app: The payee shows the QR-code that encodes his address (see Figure~\ref{fig:paytm}) and the payer scans it and then send the money. Therefore, it is just as vulnerable to our attack as the Bitcoin Wallet one. 

Postepay~\cite{postepay}, the app of the Italian postal and financial services provider, is another mobile application with more than 5M installs that feature the same scan-and-pay technique through QR-code for person-to-person payments (see Figure~\ref{fig:postepay}). We modify Malview and implement another proof-of-concept malware that targets Postepay. Similarly to its first version, Malview runs in the background and it monitors information in the \emph{proc} folder to track the user interaction with the foreground app~\cite{activityAttack}. Once the victim navigates to the activity that shows the QR-code with his address the malware adds an overlay with the address of the attacker. Postepay app allows the user to also store her QR-code in the photo gallery for simplicity in the future. We find this method insecure as any malicious apps with access to users' photos can overwrite the QR-code with a fake one. Differently from Bitcoin apps, the Postepay app shows a summary page to the payer that contains the nickname of the payee, e.g., m.rossi. At this point the payer should make sure that the nickname corresponds to the real payer. But, if the attacker sets his nickname using simple \emph{visually deceptive text} techniques---the malicious nickname is similar to the one of the victim---there is a probability for the payer to miss the difference as shown in other works~\cite{phishingWarningsChi08}. 

\subsection{Virtual Business Cards}
\label{sec:vcards}
Even though the capacity of a QR-code to store data is limited to a few Kilobits, it is still an optimal approach on some particular scenarios like virtual business cards (vCards), that store all contact information of a professional. vCards can be embedded in QR-codes so to easily create a phone contact entry by just reading the QR-code. The code can be printed on a paper business card or it can simply be shown on the screen of a smartphone. App stores for both Android and iOS offer many free applications
that decode different types of QR-codes, including vCards.
However, the integrity of the QR-code can be compromised when scanned from a smartphone screen if a malware is running on the system. 
As shown our Bitcoin Wallet application attack, the malware can place a malicious QR-code on top of the original one. Although the type of data stored on the QR-code is limited to contact details, an adversary can use this attack as a starting point for more dangerous threats. For instance, the contact details can include a URL to a malicious site that tricks the user to enter credentials or download a malware. A targeted attack would help the adversary to impersonate the victim to others, i.e., commit identity theft.

\subsection{Denial of Service}
\label{sec:dos}
On some particular scenarios Malview cannot simply replace the original QR-code with a bogus one and go undetected. For example, one bank application can allow its users to exchange money with each other through scan-and-pay technique only if both parties are registered customers of the bank. From the technical perspective Malview can attack this application but the adversary risks to be easily exposed. If the fraudulent transaction is noticed either by the payer or the payee, the bank has the required information to track the adversary. Therefore, in this scenario the adversary is not interested on stealing the money directly. However, sometimes it is beneficial to cause denial of service to his victims by placing an invalid QR-code on the screen, or the QR-code of an innocent third party to cause confusion. In case the malware is installed on the smartphone of a cashier in a shop it would cause problems by preventing customers to pay as usual.

\section{Countermeasures}
\label{sec:countermeasures}
The attack presented in this paper has a direct financial impact on the victims as it steals money from users and forwards it to a malicious wallet. Although the attack technique is similar to UI redressing, it has several characteristics that make possible defense mechanisms challenging.
In this section we discuss possible approaches that can serve as countermeasures to our attack.

\subsection{Using the Touch Filtering Specific in Android} 
\label{sec:touchFiltering}
Android provides applications with a defense mechanism to prevent clickjacking attacks. The idea of this protection is to preserve the integrity of the application interface when the user clicks a particular button. The touch filtering mechanism is disabled by default, but developers can enable it for a given view object. When enabled for a specific view, the system will discard all clicks (touch events) issued over areas of the view obscured by another service's window, thus the view will not get any touch event. 
Therefore, this mechanism thwarts only attacks that make use of ``passive'' overlays---visible labels or images to the user, but do not get touch events~\cite{clickjackingBlackhat12, vigna15}. 
However, the touch filtering based mechanism is useless in our attack where the overlay is active and it prevents the foreground application from receiving touch events.

\begin{figure}
\centering
\includegraphics[width=0.18\textwidth, keepaspectratio]{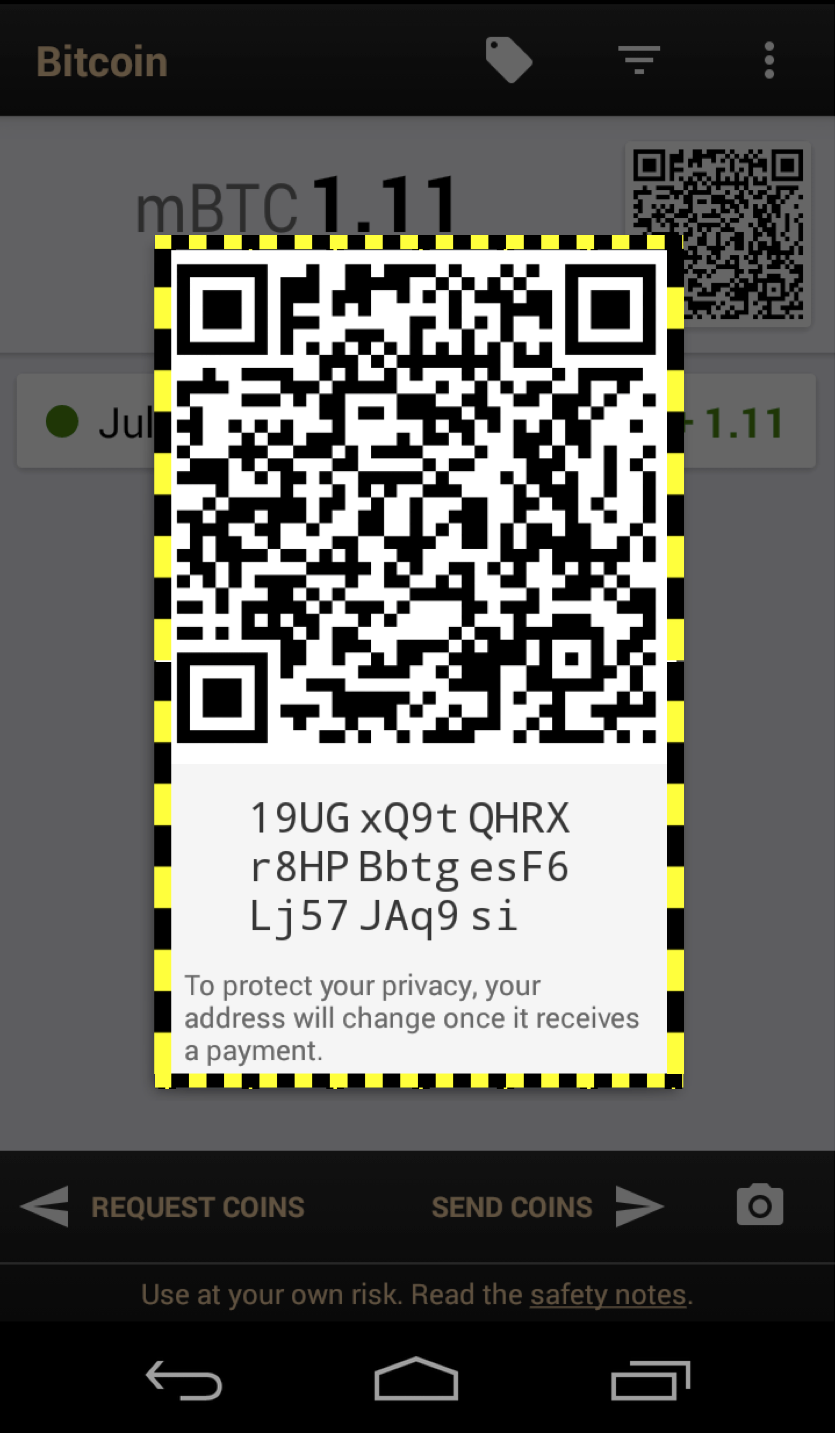}
\caption{A screenshot of the attack countermeasure: The system-level framing with yellow and black stripes of alert windows.}
\label{fig:border_defense}
\end{figure}

\subsection{Making Background service Views Easy to Recognize} 
\label{sec:border_countermeasure}
One possible way to mitigate the attack is to modify, at the Android OS system level, the appearance of views generated by a background service so to make them easily recognizable by the user by restricting their styling. We modified the Android OS so to impose all such view components to have a well-distinguishable frame-box or texture pattern, or both. In addition, the system also enforces overlay views adhere to this specific style, and forbid it for any other view type. This way, even transparent views like the ones Malview makes use of become recognisable by the user, as shown in Figure~\ref{fig:border_defense}.

To evaluate the effectiveness of this defense mechanism we recruited 90 participants for a user study in a university campus and organized an in-person interview (see Table~\ref{tab:border_experiment}). The participants were presented with a Samsung Galaxy S5 device running our modified version of the Android OS. We have installed the Malview malware on the device, running in the background. Participants were guided to open the Bitcoin Wallet application and play the role of a receiver in the transaction. When the QR-code is loaded, users see the framebox imposed by the system as in Figure~\ref{fig:border_defense}.

In the end of the interviews we asked participants for their comments regarding the black and yellow stripes around the QR-code. We learned that it was not straightforward for all the users to relate this to security concerns. The user study shows that 36\% of the participants did not get alerted of an ongoing attack even with the countermeasure in place, as demonstrated in Figure~\ref{fig:border_defense}. This result shows that if such countermeasure is implemented in Android, users should be trained to understand how overlays work and therefore how to be protected.

\textbf{Ethical Considerations.} The experiments were carried out by lending to each of the volunteers our own devices. At no point did we require participants to use their own devices or provide any private or sensitive information like usernames, passwords or private keys.

\begin{table}[t]
	\centering
	\begin{footnotesize}
			\caption{Demographics of experiment participants.}
	\begin{tabular}{c  c  c | c  c | c  c | c}
		\toprule
		\multicolumn{3}{c |}{\textbf{Age}} & \multicolumn{2}{c |}{\textbf{Gender}} & \multicolumn{2}{c |}{\textbf{Android users}} & {\textbf{Total}}\\
		20-25 & 25-30 & 30-35 & M & F & Yes & No & {}\\
		\midrule
		78 & 10 & 2 & 49 & 41 & 50 & 40 & 90\\
		\bottomrule
	\end{tabular}
	\label{tab:border_experiment}
		\end{footnotesize}
\end{table}

\subsection{One-time Codes}
Some mobile payment applications require the sender to type a one-time password (OTP) to confirm the transaction. For instance, Paytm~\cite{paytm} sends a text message with an OTP code to the payer's phone number.
This solution harms the goals of Bitcoin and does not offer the proper protection. Bitcoin is a decentralized system and transactions should not depend on a single or limited number of entities such as telecom networks. In addition, to deliver the OTP through SMS the network should be able to link the sender's wallet with the respective phone number.
However, the OTP does not protect against the presented attack because the code is not descriptive of the transaction and the payer would type it normally thinking she is confirming the transaction to the genuine receiver.

\subsection{Sensitive Views}
Previous works~\cite{hoverPaper, clickShield} present the concept of ``sensitive'' views that allow developers to protect a specific component of their application. This feature protects legitimate applications against UI redressing attacks because it forces any overlay created by another application (e.g., a background service) to move outside of the area of the sensitive view. This defense scheme mitigates the attack and it does neither change the user experience, nor place the burden to the user. However, application developers are responsible to mark properly the sensitive views of their wallet applications that show the QR-codes in order to inform the system that particular components should be protected. Then, the system would prevent any malicious service trying to load an overlay with a counterfeit QR-code representing the address of the adversary on top of the original one. Furthermore, mobile applications are typically updated to include new features or adapt the application to the new hardware, therefore, even if new UI redressing attacks are seen in the wild, the developers have the opportunity to increase their protection through sensitive views. This approach can be easily integrated in the current applications serving as wallets of cryptocurrencies or other mobile payments based on the scan-and-pay (e.g., QR-codes) technology.

\section{Conclusions}
\label{sec:conclusion}
In this paper we show how the scan-and-pay method can be exploited by malicious applications to steal the transaction ammount in mobile payments of e-coins. We target Bitcoin wallet applications that support payments based on QR-codes. The goal of the attack is to steal transaction coins directed to the payee's wallet when the QR-codes is scanned. We exploit features of the Android OS that allow any background service to show overlays on top of the foreground application. Our prototype application makes use of these overlays to achieve the goal of stealing the transaction amount in a stealthy way---without nor the payer neither the payee understanding what is actually going on. Our experiments show that the attack is fast and efficient. Lastly, we discuss other use-cases of the attack and possible countermeasures to prevent it.

\bibliographystyle{IEEEtranS}
\bibliography{bib}

\begin{thebibliography}{10}
\providecommand{\url}[1]{#1}
\csname url@samestyle\endcsname
\providecommand{\newblock}{\relax}
\providecommand{\bibinfo}[2]{#2}
\providecommand{\BIBentrySTDinterwordspacing}{\spaceskip=0pt\relax}
\providecommand{\BIBentryALTinterwordstretchfactor}{4}
\providecommand{\BIBentryALTinterwordspacing}{\spaceskip=\fontdimen2\font plus
\BIBentryALTinterwordstretchfactor\fontdimen3\font minus
  \fontdimen4\font\relax}
\providecommand{\BIBforeignlanguage}[2]{{%
\expandafter\ifx\csname l@#1\endcsname\relax
\typeout{** WARNING: IEEEtranS.bst: No hyphenation pattern has been}%
\typeout{** loaded for the language `#1'. Using the pattern for}%
\typeout{** the default language instead.}%
\else
\language=\csname l@#1\endcsname
\fi
#2}}
\providecommand{\BIBdecl}{\relax}
\BIBdecl

\bibitem{clickjackingUSENIX14}
D.~Akhawe, W.~He, Z.~Li, R.~Moazzezi, and D.~Song, ``Clickjacking revisited: A
  perceptual view of {UI} security,'' in \emph{Proc. of USENIX WOOT}, 2014.

\bibitem{maintainUI}
A.~AlJarrah and M.~Shehab, ``Maintaining user interface integrity on android,''
  in \emph{Proc. of IEEE COMPSAC}, 2016.

\bibitem{androidDev}
{Android Developers}, ``{Android Developers},''
  \url{https://developer.android.com}, accessed jan 2019.

\bibitem{instantApps}
S.~Aonzo, A.~Merlo, G.~Tavella, and Y.~Fratantonio, ``Phishing attacks on
  modern android,'' in \emph{Proceedings of the 2018 ACM SIGSAC Conference on
  Computer and Communications Security}, ser. CCS '18, 2018.

\bibitem{vigna15}
A.~Bianchi, J.~Corbetta, L.~Invernizzi, Y.~Fratantonio, C.~Kruegel, and
  G.~Vigna, ``{What the App is That? Deception and Countermeasures in the
  Android User Interface},'' in \emph{{Proc. of IEEE S\&P}}, 2015.

\bibitem{walletSource}
{Bitcoin Wallet developers}, ``{Bitcoin Wallet developers - Github},''
  \url{https://github.com/bitcoin-wallet}, accessed nov 2018.

\bibitem{activityAttack}
Q.~A. Chen, Z.~Qian, and Z.~M. Mao, ``Peeking into your app without actually
  seeing it: Ui state inference and novel android attacks,'' in \emph{Proc. of
  USENIX Security}, 2014.

\bibitem{phishingWarningsChi08}
S.~Egelman, L.~F. Cranor, and J.~Hong, ``{You've been warned: an empirical
  study of the effectiveness of web browser phishing warnings},'' in
  \emph{Proc. of SIGCHI CHI}, 2008.

\bibitem{fernandesFC}
E.~Fernandes, Q.~A. Chen, J.~Paupore, G.~Essl, J.~A. Halderman, Z.~M. Mao, and
  A.~Prakash, ``{Android UI Deception Revisited: Attacks and Defenses},'' in
  \emph{Proc. of FC}, 2016.

\bibitem{cloackDagger}
Y.~Fratantonio, C.~Qian, S.~P. Chung, and W.~Lee, ``Cloak and dagger: From two
  permissions to complete control of the ui feedback loop,'' in \emph{Proc. of
  IEEE S\&P}, 2017.

\bibitem{indiaCrise}
{G. Anand and H. Kumar}, ``{India Hobbles Through a Cash Crisis, and Electronic
  Payments Boom - The New York Times},'' \url{https://nyti.ms/2FAUq3z},
  accessed oct 2018.

\bibitem{bitcoinWallet}
{Google Play}, ``{Bitcoin Wallet},'' \url{http://bit.ly/2QSWsy7}, accessed oct
  2018.

\bibitem{forceReboot}
H.~Huang, S.~Zhu, K.~Chen, and P.~Liu, ``From system services freezing to
  system server shutdown in android: All you need is a loop in an app,'' in
  \emph{Proc. of ACM CCS}, 2015.

\bibitem{repackage2}
H.~Huang, S.~Zhu, P.~Liu, and D.~Wu, \emph{A Framework for Evaluating Mobile
  App Repackaging Detection Algorithms}.\hskip 1em plus 0.5em minus 0.4em\relax
  Berlin, Heidelberg: Springer Berlin Heidelberg, 2013, pp. 169--186.

\bibitem{clickjackingUSENIX12}
L.~Huang, A.~Moshchuk, H.~J. Wang, S.~Schechter, and C.~Jackson,
  ``Clickjacking: attacks and defenses,'' in \emph{Proc. of USENIX Security},
  2012.

\bibitem{storecrawler}
{IzzyOnDroid}, \url{http://android.izzysoft.de/intro.php}, accessed jun 2018.

\bibitem{btcDouble}
O.~G. Karame, E.~Androulaki, and S.~Capkun, ``Double-spending fast payments in
  bitcoin,'' in \emph{Proc. of ACM CCS}, 2012.

\bibitem{tapprints2012}
E.~Miluzzo, A.~Varshavsky, S.~Balakrishnan, and R.~R. Choudhury, ``{Tapprints:
  Your Finger Taps Have Fingerprints},'' in \emph{Proc. of ACM MobiSys}, 2012.

\bibitem{satoshiBitcoin}
S.~Nakamoto, ``{Bitcoin: A Peer-to-Peer Electronic Cash System},'' Nov. 2008,
  \url{https://bitcoin.org/bitcoin.pdf}.

\bibitem{clickjackingBlackhat12}
M.~Niemietz and J.~Schwenk, ``Ui redressing attacks on android devices,'' in
  \emph{Black Hat, Abu Dhabi}, 2012.

\bibitem{paytm}
{Paytm - One97 Communications Ltd.}, ``{Mobile Recharge, DTH, Bill Payment,
  Money Transfer},'' \url{https://bit.ly/2QW1pX2}, accessed oct 2018.

\bibitem{clickShield}
A.~Possemato, A.~Lanzi, S.~P.~H. Chung, W.~Lee, and Y.~Fratantonio,
  ``Clickshield: Are you hiding something? towards eradicating clickjacking on
  android,'' in \emph{Proc. of ACM CCS}, 2018.

\bibitem{postepay}
{Poste Italiane S.p.A.}, ``{PostePay - Android Apps on Google Play},''
  \url{https://bit.ly/2FEh4Z4}, accessed oct 2018.

\bibitem{windowGuard}
C.~Ren, P.~Liu, and S.~Zhu, ``Windowguard: Systematic protection of gui
  security in android,'' in \emph{Proc. of NDSS}, 2017.

\bibitem{taskJacking15}
C.~Ren, Y.~Zhang, H.~Xue, T.~Wei, and P.~Liu, ``{Towards Discovering and
  Understanding Task Hijacking in Android},'' in \emph{Proc. of USENIX
  Security}, 2015.

\bibitem{clickjackingSAC15}
H.~Shahriar and H.~Haddad, ``Security assessment of clickjacking risks in web
  applications: metrics based approach,'' in \emph{Proc. of ACM SAC}, 2015.

\bibitem{hoverPaper}
E.~Ulqinaku, L.~Malisa, J.~Stefa, A.~Mei, and S.~\v{C}apkun, ``{Using Hover to
  Compromise the Confidentiality of User Input on Android},'' in \emph{Proc. of
  ACM WiSec}, 2017.

\bibitem{taplogger2012}
Z.~Xu, K.~Bai, and S.~Zhu, ``Taplogger: Inferring user inputs on smartphone
  touchscreens using on-board motion sensors,'' in \emph{Proc. of ACM WISEC},
  2012.

\bibitem{zhouIdentity13}
X.~Zhou, S.~Demetriou, D.~He, M.~Naveed, X.~Pan, X.~Wang, C.~A. Gunter, and
  K.~Nahrstedt, ``Identity, location, disease and more: inferring your secrets
  from android public resources,'' in \emph{Proc. of ACM CCS}, 2013.

\end{thebibliography}

\end{document}